\documentclass[preprint,showkeys,showpacs,preprintnumbers,amsmath,amssymb]{revtex4}
\usepackage{graphicx}
\usepackage{dcolumn}	
\usepackage{bm}	       

\usepackage{amssymb, amsmath}
\usepackage{mathrsfs}  

\newcommand{\D}{\partial}
 \newcommand{\ve}[1]{{\bf #1}}
 \newcommand{\vve}[1]{{\bf { #1}}}
 \newcommand{\veg}[1]{{\boldsymbol {#1}}}	
 \newcommand{\vveg}[1]{{\boldsymbol {#1}}}	

 \newcommand{\const} {{\rm const}}

 \newcommand{\delt} {{\Delta t}}
 \newcommand{\mean}[1]{{\overline {#1}}}

\newcommand{\mat}[4]{ \bracket{\begin{array}{ll}
				#1  &#2\\
				#3  &#4
				\end{array}}}

\newcommand{\matt}[9]{ \bracket{\begin{array}{lll}
				#1  &#2  &#3\\
				#4  &#5  &#6\\
				#7  &#8  &#9
				\end{array}}}
\newcommand{\vectt}[3]{ \bracket{\begin{array}{l}
				#1 \\
				#2 \\
				#3
				\end{array}}}

\newcommand{\DD}[2]{\frac {\D #1}{\D #2}}

\newcommand{\dt}[1]{\frac {d#1} {d t}}

\newcommand{\bracket}[1]{\left[#1\right]}

\newcommand{\parenth}[1]{\left(#1\right)}

\DeclareMathSymbol{\R}{\mathbin}{AMSb}{"52}

 \newcommand{\info}{{T}}
 \newcommand{\Nino}{{Ni\~no}}
 \newcommand{\ENSO}{{El~Ni\~no} }

 \newcommand{\DI}[1]{\frac {\D #1} {\D x_1}}
 
 \newcommand{\DIDI}[1]{\frac {\D^2 #1} {\D x_1^2}}

\begin{document}

   \title{\bf 
	Causality between time series\footnote
	{This material was presented at the 18th Conference on 
	Atmospheric and Oceanic Fluid Dynamics, 13-17 June 2011, 
	Spokane, WA, under the title {\it Information Flow and
	Causality within Atmosphere-Ocean Systems}.
 	PPT slides are available from the online conference archives.}}

        \author{X. San Liang\footnote
		{URL: http://people.deas.harvard.edu/$\sim$san/ }}

	\email{sanliang@courant.nyu.edu}
	\affiliation{Nanjing Institute of Meteorology, Nanjing 210044, and\\
	China Institute for Advanced Study,\\
	Central University of Finance and Economics, Beijing 100081, China}


\begin{abstract}
{
Given two time series, can one tell, in a rigorous and quantitative way, 
the cause and effect between them?  Based on a recently rigorized 
physical notion namely information flow, we arrive at a concise formula and 
give this challenging question, which is of wide concern in different 
disciplines, a positive answer. Here causality is measured by the time rate 
of change of information flowing from one series, say, $X_2$, to another, 
$X_1$. The measure is asymmetric between the two parties and, particularly, 
if the process underlying $X_1$ does not depend on $X_2$, then the resulting 
causality from $X_2$ to $X_1$ vanishes. 
The formula is tight in form, involving only the commonly used statistics, 
sample covariances. It has been validated with touchstone series 
purportedly generated with one-way causality. It has also been applied 
to the investigation of real world problems; an example presented here 
is the cause-effect relation between two climate modes, 
El Ni\~no and Indian Ocean Dipole, which have been linked to the hazards 
in far flung regions of the globe, with important results that would
otherwise be difficult, if not impossible, to obtain.

}
\end{abstract}



\maketitle

\section{Introduction}
Causality identification between dynamical events is a subject of great
interest in different disciplines. Examples are seen everywhere in
neuroscience\cite{Pereda}\cite{Korz03}\cite{Wu}\cite{Baptista},
biology\cite{Sachs}\cite{Okasha}\cite{Jost}, 
social and computer network dynamics\cite{Ay}\cite{Sommerlade},
financial economics\cite{ChenCR}\cite{Lee}, 
statistical physics\cite{Crutchfield},
to name a few. 
Toward the end of this study, we will see a climate science problem 
that has been linked to the severe weather
and natural disasters on a global scale.

Given that, more often than not, what is known about the two events in
question are in the form of time series, causality analysis
between time series is therefore of particular importance. 
Presently a common practice in applied sciences, e.g., 
climate science, is through computing time-lagged 
correlation. However, it is well known that correlation does not
carry the needed directedness or asymmetry and hence does not necessarily
imply causality. Besides, it is easy to argue that, for recurrent processes, 
there is actually no way to distinguish a lag from an advance,
unless one has enough {\it a priori} knowledge of the processes of concern. 
This is particularly a problem when the processes are nonsequential (e.g.,
those in the nonsequential stochastic control systems).
    %
    %
Another common practice is through Granger causality test, which 
is a statistical test of the usefulness of one time series in forecasting 
another. This kind of test, as the name implies, provides only a yes/no
judgment, lacking the quantitative information that may be required in 
many circumstances.

On the other hand, information flow, or information transfer as it 
may be referred to in the literature, provides 
such a quantitative measure; the amount of information exchange between two
events offers not only the direction but also the magnitude of the
cause-effect relation. Due to its importance, the past decades have seen a
surge of interest in formulating this notion. Formalisms
have been established empirically or half-empirically, 
and have been used for different researches,
    among which is transfer entropy\cite{Schreiber}.
Realizing that information flow is a real physical notion, and that a real 
physical notion should be rigorously, rather than empirically,
built on a solid foundation 
so as to be universally applicable for problems in different disciplines,
Liang and Kleeman\cite{LK05} take the initiative to establish, 
followed by a series of researches afterwards, a rigorous formalism 
for the information flow within given dynamical systems, 
both deterministic and stochastic systems (see \cite{Liang13} for a review).
The problem now is whether and how the same notion can be translated to
cases with no dynamics but time series previously given; 
that way, the quantitative causal relation between the series 
will be obtained. In some sense this is an inverse problem, and could be
challenging since no {\it a priori} knowledge of the dynamics is available.
In the following we first briefly introduce the Liang-Kleeman information
flow within a two-dimensional system (section~\ref{sect:LK}). The major
derivation is presented in section~\ref{sect:causality}, where a concise
formula for causality analysis is obtained. This formula is 
validated with touchstone time series purportedly generated with only
one-way causality (section~\ref{sect:validation}); it is then applied to
the study of the causal relation between \ENSO and Indian Ocean Dipole (IOD), 
the two major modes of climate variation, with important results
which have been evidenced for years but missed in previous
analyses with existing tools (section~\ref{sect:application}).

\section{Liang-Kleeman information flow}	\label{sect:LK} 

Let us begin with a brief introduction of the Liang-Kleeman information
flow. Consider a $d$-dimensional stochastic system
	\begin{eqnarray}	\label{eq:gov}
	d\ve X = \ve F(\ve X; \veg\theta) dt + \vve B(\ve X; \veg\theta) d\ve W
	\end{eqnarray}
where $\ve F$ is the vector of drift coefficients, 
$\vve B = (b_{ij})$ a matrix of diffusion coefficients 
(or volatility in finance), 
and $\ve W$ a vector of standard Wiener process ($\dot {\ve W}$ is the
so-called white noise). The rate of information flowing 
from a component, say, $X_2$ 
to another, say, $X_1$, is the change rate of the marginal entropy of
$X_1$, minus the same change rate but with the effect from $X_2$ 
instantaneously excluded from the system. 
These rates of information flow/transfer 
have been obtained analytically in a closed form for any given dynamical
systems\cite{Liang08}; particularly, for a system of dimensionality 2,
which we will be considering in this study, the flow rate 
from $X_2$ to $X_1$ is
	\begin{eqnarray}	\label{eq:transfer1}
	\info_{2\to1} = - E \parenth{\frac 1 {\rho_1} \DI {(F_1\rho_1)}}
	    + \frac 1 2 E \parenth{\frac 1 {\rho_1} 
			\DIDI{(b_{11}^2+b_{12}^2)\rho_1}},
	\end{eqnarray}
where $\rho_1$ is the marginal probability density of $X_1$, and
$E$ the mathematical expectation.
$\info_{2\to1}$ may be either zero or nonzero. A nonzero $\info_{2\to1}$
means that $X_2$ is causal to $X_1$: a positive value means that $X_2$
makes $X_1$ more uncertain, and vice versa.
This measure of information flow possesses an important property, 
namely the property of flow/transfer asymmetry: 
One-way information flow implies nothing about the 
transfer in the opposition direction. A very particular case
is: when the evolution of $X_1$ does not depend on $X_2$, $\info_{2\to1}=0$
(see \cite{Liang13} for proofs in different situations). 
Considering its criticality, this property makes a
touchstone for the testing of causality analysis tools.

\section{Causality analysis}	\label{sect:causality}

Shown above are some of the theoretical results of the Liang-Kleeman 
information flow for 2D systems. 
These results  are rigorous, but the formalism is not about causality analysis, 
as the causal relation is prescribed in the given system.
We need to develop an analysis without {\it a priori} knowledge of the
dynamics; the only given conditions are the two time series. To some extent 
this is like an inverse problem. Generally it could be very challenging, 
considering the touchstone property, among others, to be verified. 

Since the dynamics is unknown, we need to choose a model first.
As always, a linear model is the natural choice.
In Eq.~(\ref{eq:gov}), assume $\ve F = \ve f + \vve A \ve X$, with
$\ve f = ({f_1}, {f_2})^T$, $\vve A = (a_{ij})$, and $\vve B = (b_{ij})$
being constant vector/matrices.
In the equation, suppose that initially $(X_1, X_2)$ has a bivariate 
normal distribution. As the system is linear, then the density is always
normal\cite{Lasota}. Denote the mean by $\veg\mu = (\mu_1, \mu_2)^T$, 
and the covariance matrix by $\veg\Sigma=(\sigma_{ij})$. They evolve
following the differential equations
	\begin{eqnarray}
	&& \dt{\veg\mu} = \ve f + \vve A \veg\mu,	\label{eq:mean}\\	
	&& \dt{\vveg \Sigma} = \vve A \vveg\Sigma + \vveg\Sigma \vve A^T 
		+ \vve B\vve B^T.			\label{eq:cov}
	\end{eqnarray}
The information flows between $X_1$ and $X_2$ are now easy to evaluate.
Consider $\info_{2\to1}$ first.
When $(b_{ij})$ are constant, the last term of (\ref{eq:transfer1})
vanishes, thanks to a nice property proved in \cite{Liang08}. 
Substitution of 
	\begin{eqnarray}
	&&F_1 = f_1 + a_{11} X_1 + a_{12} X_2			\\
	&&\rho_1 = \frac 1 {\sqrt{2\pi} \sigma_1} 
		   e^{-\frac {(x_1-\mu_1)^2} {2\sigma_1^2}}
	\end{eqnarray}
into (\ref{eq:transfer1}) for $F_1$ and $\rho_1$ yields,
after some algebraic manipulations,
	\begin{eqnarray}
	\info_{2\to1} 
	&=& \frac {\sigma_{12}} {\sigma_{11}} a_{12},	\label{eq:info21}
	\end{eqnarray}
where $\sigma_{ij}$ is obtained by solving (\ref{eq:cov}). (Note that
extra terms will be involved if $\vve B$ depends on $\ve X$.)
Likewise,
	\begin{eqnarray}
	\info_{1\to2} 
	   &=& \frac {\sigma_{12}} {\sigma_{22}} a_{21}. \label{eq:info12}
	\end{eqnarray}

\begin{figure}[h]
	\begin{center}
	\includegraphics[angle=0, width=0.5\textwidth]
		{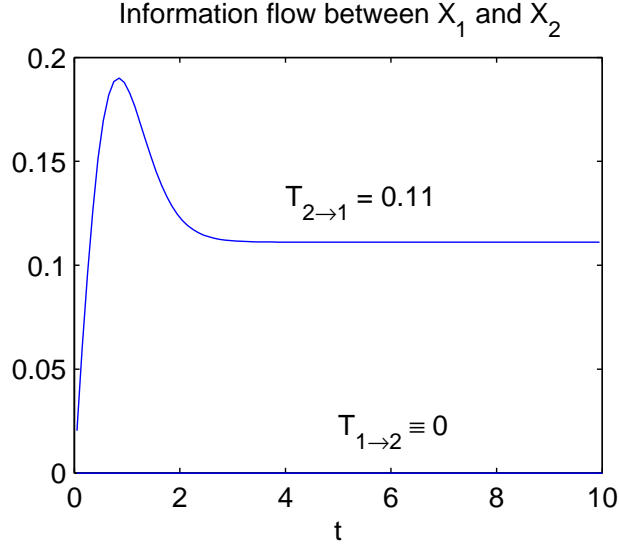}
	\caption{Accurate result of the information flow
	between $X_1$ and $X_2$ within the linear system 
	(\ref{eq:ex1})-(\ref{eq:ex2}).
	The variance is initialized with $\sigma_{11}=\sigma_{22}=0.1$,
	$\sigma_{12}=0$. \protect{\label{fig:info_anal}}}
	\end{center}
	\end{figure}

For an example, consider the case 
	$\ve f = \ve 0$,
	$\vve A = \mat {-1} {0.5} {0} {-1}$,
	$\vve B = \mat {0.1} 0 0 {0.1}$,
which gives a 2D stochastic differential equation (SDE):
	\begin{eqnarray}
	&& dX_1 = (-X_1 + 0.5 X_2) dt + 0.1 dW_1, 	\label{eq:ex1}\\
	&& dX_2 = -X_2 dt + 0.1 dW_2.			\label{eq:ex2}
	\end{eqnarray}
Clearly, $X_2$ drives $X_1$, but not vice versa.
The computed rates of information flow are shown in
Fig.~\ref{fig:info_anal}. As expected, $T_{1\to2}\equiv0$, since the
evolution of $X_2$ does not depend on $X_1$. That is to say, to $X_2$, 
$X_1$ is not causal. On the other hand, $X_2$ drives $X_1$ and hence is
causal to $X_1$; correspondingly $T_{2\to1} \ne 0$.
In this example, $T_{2\to1}$ approaches a constant 0.1111 
no matter how the system
is initialized, though the result may be different during the spin-up
period. This kind of example is very typical in causality analysis:
One component causes another, but the latter has no feedback to the
former. Later we will use it to test our causality analysis.
	
Our strategy to approach the problem is first to estimate the linear model, 
i.e., to estimate the parameters $(\ve f, \vve A, \vve B)$, which we will
be symbolically writing as $\veg\theta$, with the time series, 
and then compute the information flow. To further simplify the
problem, let $\vve B = \mat {b_1} 0 0 {b_2},$ and assume that the time series
are equal-distanced. 

We use maximum likelihood estimation (mle) to fulfill the
purpose. Consider an interval $[n\delt, (n+1)\delt]$, $\delt$ being the
time stepsize. If the transition 
probability density function $\rho(\ve X_{n+1} | \ve X_n; \veg\theta)$ 
can be obtained,
since $\{X_n\}$ is a Markov process, the likelihood
	\begin{eqnarray}
	L_N(\veg\theta) = \rho(\ve X_1) \cdot \rho(\ve X_2 | \ve X_1;
	\veg\theta) ... \rho(\ve X_N | \ve X_{N-1}; \veg\theta)
	\end{eqnarray}
is then given ($N$ the sample size), or, alternatively, the log likelihood
	\begin{eqnarray}
	\ell_N(\veg\theta) = \sum_{n=1}^N \log \rho(\ve X_{n+1} | \ve X_n;
		\veg\theta) + \log\rho(\ve X_1)
	\end{eqnarray}
is obtained. When $N$ is large, $\rho(\ve X_1)$ can be dropped without
causing much error. In this linear case, analytical solution of 
$\rho$ can be found; the result, however, is complicated. In the interest 
of applicability, we turn to the discretized version of the SDE. 
Using the Euler-Bernstein scheme (cf.~\cite{Lasota}),
	\begin{eqnarray}
	\ve X_{n+1} = \ve X_n + \ve F(\ve X_n; \veg\theta) \delt
	+ \vve B(\ve X_n; \veg\theta) \Delta \ve W,
	\end{eqnarray}
where $F=\ve f + \vve A \ve x$. Suppose $\delt$ is small, we know 
	\begin{eqnarray*}
	&& \Delta \ve W\ \sim\ {\mathscr N}(\ve 0,\ \vve B \vve B^T \delt), \\
	&& \ve X_{n+1} | \ve X_n = \ve x_n\ \sim\ 
	{\mathscr N} (\ve x_n + \ve F\delt,\ \vve B\vve B^T\delt).
	\end{eqnarray*}
So
	\begin{eqnarray*}
	&& \rho(\ve X_{n+1} =\ve x_{n+1}| \ve X_n=\ve x_n)	\cr
	&& = \frac 1 {\bracket{(2\pi)^2 \det(\vve B\vve B^T\delt)}^{1/2}}
	     e^{-\frac12 (\ve x_{n+1}-\ve x_n-\ve F\delt)^T 
		  (\vve B\vve B^T\delt)^{-1} 
		  (\ve x_{n+1}-\ve x_n-\ve F\delt)},
	\end{eqnarray*}
and
   \begin{eqnarray}	
   && \ell_N(\veg\theta) 
      = \sum_{n=1}^N \log\rho(\ve X_{n+1} | \ve X_n;\veg\theta)	\cr
   && = \const - \frac12 \sum_{n=1}^N \log \bracket{\det(\vve B\vve B^T)} \cr
   &&\ \ \   -\frac1{2\delt} \sum_{n=1}^N 
	(\ve x_{n+1}-\ve x_n-\ve F\delt)^T 
	  (\vve B\vve B^T)^{-1} 
	  (\ve x_{n+1}-\ve x_n-\ve F\delt).	\label{eq:loglike}
	\end{eqnarray}	
Note here we have assumed that $N$ is large enough such that
$\rho(X_1)$ can be dropped without causing much error. 
The mle of $\veg\theta$ is, therefore,
	\begin{eqnarray}
	\veg{\hat\theta} = \arg \max_{\theta} \ell_N(\veg\theta).
	\end{eqnarray}

For convenience, denote, for $i=1,2$,
	\begin{eqnarray}
	\dot X_{i,n} &:=& \frac {X_{i,n+1} - X_{i,n}} \delt,
		\label{eq:xdot}	\\
	R_{i,n} &:=& \dot X_{i,n} - (f_i + a_{i1} X_{1,n} + a_{i2} X_{2,n}).
	\end{eqnarray}
Also notice that $\vve B \vve B^T = \mat {b_1^2} 0 0 {b_2^2}$. 
Substituting into Eq.~(\ref{eq:loglike}), the log likelihood 
$\ell_N(\veg\theta) = \ell_N(\ve f, \vve A, \vve B)$ is, 
	\begin{eqnarray}
	\ell_N(\ve f, \vve A, \vve B) = \const - \frac N 2 \log b_1^2 b_2^2
	- \frac \delt 2 \bracket{\frac 1 {b_1^2} \sum_{n=1}^N R_{1,n}^2
	                   + \frac 1 {b_2^2} \sum_{n=1}^N R_{2,n}^2}
	\end{eqnarray}
The estimators of $\ve f$, $\vve A$, and $\vve B$ can be found by
maximizing $\ell_N$. It is interesting to note that the equations
governing the estimators
$(\hat f_1, \hat a_{11}, \hat a_{12})$,
$(\hat f_2, \hat a_{21}, \hat a_{22})$,
and $(\hat b_1, \hat b_2)$ are actually decoupled.
This makes the estimation much easier. Besides, notice that
maximizing $\ell_N$ over $(f_i, a_{i1}, a_{i2})$ is equivalent to
minimizing $\sum_n R_{i,n}^2 := Q_{N,i}$ over the same group of 
parameters, for $i=1,2$. So $(\hat f_1, \hat a_{11}, \hat a_{12})$ 
are precisely the least square estimators, which satisfies
	\begin{eqnarray*}
	\matt N              {\sum X_{1,n}}         {\sum X_{2,n}}
	      {\sum X_{1,n}} {\sum X_{1,n}^2}       {\sum X_{1,n} X_{2,n}}
	      {\sum X_{2,n}} {\sum X_{2,n} X_{1,n}} {\sum X_{2,n}^2} 
	\vectt {\hat f_1} {\hat a_{11}} {\hat a_{12}}
	=
	\vectt {\sum\dot X_{1,n}}  
	       {\sum X_{1,n} \dot X_{1,n}} 
	       {\sum X_{2,n} \dot X_{1,n}}.
	\end{eqnarray*}

The above equation set can be written in a more familiar and succinct form:
	\begin{eqnarray}
	\matt 1              {\mean {X_1}}         {\mean {X_2}}
	      {\mean {X_1}}  {\mean {X_1^2}}       {\mean {X_1 X_2}}
	      {\mean {X_2}}  {\mean {X_1 X_2}}      {\mean {X_2^2}}
	\vectt {\hat f_1} {\hat a_{11}} {\hat a_{12}}
	=
	\vectt {\mean {\dot X_1}}  
	       {\mean {X_1 \dot X_1}} 
	       {\mean {X_2 \dot X_1}},
	\end{eqnarray}
where, as usual, the overline signifies sample mean.
After some algebraic manipulations, it becomes:
	\begin{eqnarray*}
	\matt 1              {\mean {X_1}}         {\mean {X_2}}
      0  {\mean{X_1^2} - \bar X_1^2}  {\mean {X_1 X_2} - \bar X_1\bar X_2}
      0  {\mean {X_1 X_2} - \bar X_1\bar X_2}   {\mean {X_2^2} - \bar X_2^2}
	\vectt {\hat f_1} {\hat a_{11}} {\hat a_{12}}
	=
	\vectt {\mean {\dot X_1}}  
	       {\mean {X_1 \dot X_1} - \bar X_1 \bar {\dot X}_1} 
	       {\mean {X_2 \dot X_1} - \bar X_2 \bar {\dot X}_1}.
	\end{eqnarray*}
Let
	\begin{eqnarray}
	&&C_{ij}   :=  \mean {(X_i - \bar X_i) (X_j - \bar X_j)}, \\
	&&C_{i,dj} := \mean {(X_i - \bar X_i) (\dot X_j - \bar {\dot X}_j)},
	\end{eqnarray}
$\vve C = (C_{ij})$ being the sample covariance matrix. The equation set is
actually
	\begin{eqnarray*}
	\matt 1  {\mean {X_1}}         {\mean {X_2}}
              0  {C_{11}}  {C_{12}}
              0  {C_{12}}  {C_{22}} 
	\vectt {\hat f_1} {\hat a_{11}} {\hat a_{12}}
	=
	\vectt {\mean {\dot X_1}}  
	       {C_{1,d1}}
	       {C_{2,d1}},
	\end{eqnarray*}
which yields
	\begin{eqnarray}
	&&\hat a_{11} = \frac {C_{22} C_{1,d1} - C_{12} C_{2,d1}} 
			    {\det \vve C}, 	\\
	&&\hat a_{12} = \frac {-C_{12} C_{1,d1} + C_{11} C_{2,d1}} 
			    {\det \vve C}, 	\\
	&&\hat f_1 = \bar {\dot X}_1 - \hat a_{11} \bar X_1 
				   - \hat a_{12} \bar X_2.
	\end{eqnarray}
The other group of estimators can be obtained by switching the indices:
	\begin{eqnarray}
	&&\hat a_{22} = \frac {C_{11} C_{2,d2} - C_{12} C_{1,d2}} 
			    {\det \vve C}, 	\\
	&&\hat a_{21} = \frac {-C_{12} C_{2,d2} + C_{22} C_{1,d2}} 
			    {\det \vve C}, 	\\
	&&\hat f_2 = \bar {\dot X}_2 - \hat a_{21} \bar X_1 
				   - \hat a_{22} \bar X_2.
	\end{eqnarray}
With these results, let $\DD {\ell_N} {b_i} = 0$ to get
	\begin{eqnarray}
	\hat b_i = \sqrt{\frac {Q_{N,i} \cdot \delt} N}, 
	\end{eqnarray}
where 
	\begin{eqnarray*}
	Q_{N,i} = \sum_{n=1}^N R_{i,n}^2
	= \sum_{n=1}^N \bracket{\dot X_{i,n} - 
		(\hat f_i + \hat a_{i1} X_{1,n} + \hat a_{i2} X_{2,n})}^2.
	\end{eqnarray*}

On the other hand, as the population covariance matrix can be estimated by
the sample covariance matrix, the estimator of $\sigma_{12}/\sigma_{11}$
is $C_{12}/C_{11}$. 
Substituting these results back to 
(\ref{eq:info21}), we finally obtain the rate of
information flowing from $X_2$ to $X_1$:
	\begin{eqnarray}
	\setlength\fboxrule{0.4pt}
	\boxed{
	\info_{2\to1} = \frac {C_{12}} {C_{11}} \cdot 
			\frac {-C_{12} C_{1,d1} + C_{11} C_{2,d1}} 
			      {C_{11}C_{22} - C_{12}^2}, 
					\label{eq:caus21} 
	}
	\end{eqnarray}
where $C_{ij}$ is the sample covariance between $X_i$ and $X_j$, and 
$C_{i,dj}$ the covariance between $X_i$ and $\dot X_j$. 
Strictly speaking, here $T_{2\to1}$ should bear a caret since it is an
estimator of the true information flow. But we will abuse the notation a
little bit for the sake of terseness. 
The flow in the opposite direction, $T_{1\to2}$,
can be directly written out by switching the indices 1 and 2.
The units are in nats per unit time.

Given a significance level,
we may estimate the confidence interval for (\ref{eq:caus21}).
This can always be achieved with bootstrap. But here things can be simplified.
When $N$ is large, $T_{2\to1}$ is approximately normally distributed
around its true value with a variance 
$\parenth{\frac{C_{12}}{C_{11}}}^2 \hat \sigma^2_{a_{12}} $,
thanks to the mle property. Here $\hat\sigma^2_{a_{12}}$
is determined as follows. 
Denote $\veg\theta = (f_1, a_{11}, a_{12}, b_1)$. Compute
	\begin{eqnarray}
	NI_{ij} = - \sum_{n=1}^N 
		\frac{\D^2 \log\rho(\ve X_{n+1} | \ve X_n;\ \hat{\veg\theta})}
		     {\D\theta_i \D\theta_j}
	\end{eqnarray}
to form a matrix $N\vve I$, $\vve I$ being the Fisher information matrix.
The inverse $(N\vve I)^{-1}$ is the covariance matrix of $\hat{\veg\theta}$,
within which is $\hat\sigma_{a_{12}}^2$. Given a significance level, 
the confidence interval can be found accordingly.

\section{Validation}	\label{sect:validation}

\begin{figure}[h]
	\begin{center}
	\includegraphics[angle=0, width=0.5\textwidth]
		{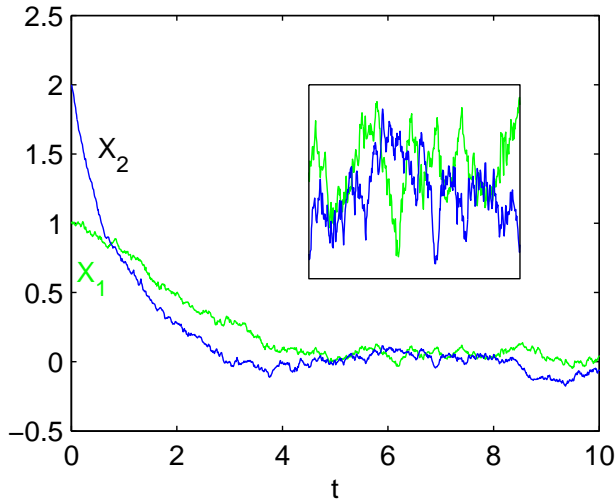}
	\caption{A sample path of the linear SDE
		(\ref{eq:ex1})-(\ref{eq:ex2}) generated using the
		Euler-Bernstein scheme with $\delt=0.001$ and
		initialized with $(1,2)$. 
		Inserted is a close-up over the interval $t=4.5$-$8.5$.
	 \protect{\label{fig:gen_series}}}
	\end{center}
	\end{figure}

The above formula is remarkably tight in form,
involving only the common statistics namely sample covariance matrix.
We now check whether they verify the example shown in
Fig.~\ref{fig:info_anal}. The data we have is just one realization, i.e.,
one sample path of $X_1$ and $X_2$ each generated from the given system.
Using a time step $\delt=0.001$, we generate 100000 steps, 
corresponding to a time span $t=0-100$. 
For later use, we purportedly initialize the series far from the
equilibrium to allow for a period of spin-down, as shown in
Fig.~\ref{fig:gen_series}. The series reach a stationary 
state after approximately $t=4$; shown in the inserted box
is a close-up plot. 
Our objective is, of course, not to estimate the whole evolution course
as shown in Fig.~\ref{fig:info_anal}; what we expect is to see whether 
the stationary values ($T_{2\to1}=0.1111$, $T_{1\to2}=0$) can be 
estimated with acceptable confidence. 
For this purpose, we form different series from the sample path
with different resolutions and different time intervals and lengths, 
and then test the performance of estimation. The testing results are 
listed in Table~\ref{tab}. 

	\begin{table}[h]
	\begin{center}
	\caption{The information flow computed with different series
	formed by sampling from the path shown in Fig.~\ref{fig:gen_series}. 
	$\Delta n$ is the sampling interval.
	The accurate result is: $T_{2\to1}=0.1111$, $T_{1\to2}=0$
	(in nats per unit time).
	In the last two cases, data on $[5,10]$ are used to compute
	$C_{ij}^*$.
	\protect{\label{tab}}}
	\begin{tabular}{ccccc}
	\hline
	\hline
	Time span & $\Delta n$ & $T_{2\to1}$ &$T_{1\to2}$ & Remark\\
	\hline
	 $t$=5-100 & 1   &  0.11 &$-2.0\times 10^{-3}$& \\
	 \	   & 20  & 0.10 &$-4.0\times 10^{-3}$ & \\
	 \	   & 100 & 0.09 &-0.01		       & \\
	\hline
	 $t$=10-20 & 1   & 0.60 & 0.17		       & \\
	 \	   & 10  & 0.57 & 0.20		       & \\
	\hline
	 $t$=0-10  & 1   & 0.74 & 0.10		       & \\
	 \	   & 1   & 0.29 & 0.02	   & use (\ref{eq:caus21a})\\
	 \	   & 10  & 0.28 & 0.02     & use (\ref{eq:caus21a})\\		
	\hline
	\end{tabular}
	\end{center}
	\end{table}

Clearly, so long as the time span of the series is long enough, the
estimation can be made rather accurate. With stationary data ($t=5-100$), 
even when one samples the path every 100 time steps (corresponding to a 
time resolution of 0.1) which yields a time series of only 1000 data points, 
the result is still acceptable to some extent. 
In some cases, the span may not be that long. To see how the formula
may perform, we make two series spanning only 10 time units. 
In the first case we test with a stationary
segment $t=10-20$. The result: 
	$T_{2\to1}=0.60$, $T_{1\to2}=0.17$,
unfortunately, is not as good as one would like to expect, 
even though the data points amount to 10000. Nonetheless,
if we estimate the standard error at a 95\% significance level, which
gives a value of 0.54 and 0.47, respectively, the result 
still sounds informative. It tells that $T_{2\to1}$ is significantly (at a
95\% level) greater than zero, while $T_{1\to2}$ is not significantly
different from zero, consistent with the accurate result. 

On the other hand, if the time span, though small, contains the 
nonstationary period, things may actually turn better. Choose an
interval $t=0-10$. The result looks remarkably promising, provided that we treat
the covariances in the first fraction of (\ref{eq:caus21}) and that in the
second fraction differently. In the first fraction, the sample covariances 
are indeed used to estimate the populate covariances, and hence cannot be
computed with the nonstationary data. In the second fraction, the mle is
for the deterministic coefficient and so nonstationary data could serve the
purpose better (indeed this is true). To distinguish these two different
estimations, we denote
the former with an asterisk; 
in doing so the formula (\ref{eq:caus21}) now looks
	\begin{eqnarray}
	\info_{2\to1} = \frac {C_{12}^*} {C_{11}^*} \cdot 
			\frac {-C_{12} C_{1,d1} + C_{11} C_{2,d1}} 
			      {C_{11}C_{22} - C_{12}^2}. 
					\label{eq:caus21a} 
	\end{eqnarray}
Now go back to the chosen time series spanning $[0,10]$.
If $C_{ij}^*$ are computed with data on $[5,10]$, 
then the one-way causality between the two series 
actually can be fairly reproduced, though the resulting
$T_{2\to1}$ is 1.6 times larger. 
Moreover, this result is very robust even the 
resolution is reduced by ten times. If one only considers a qualitative
causal relation, this example shows that our approach is promising even 
when the length of the time series is limited.

We have tested the formula against systems with different drift and 
diffusion coefficients, particularly systems where noises dominate.
Again, so long as the series have a span long enough, the causal relation
can be faithfully reproduced. As an example, for the system
	\begin{eqnarray*}
	&& dX_1 = -0.5 X_1 + X_2 + 20 dW_1, \\
	&& dX_2 = -0.7 X_2 + 10 dW_2,
	\end{eqnarray*}
if the series span 1000 time units or longer, the estimation of 
$T_{2\to1}$ and $T_{1\to2}$ can be made fairly accurate.

\section{An application example}		\label{sect:application}

Finally let us look at an application to a real world problem:
the causal relation between the two famous climate modes, 
\ENSO and Indian Ocean Dipole (IOD).
\ENSO is the strongest interannual climate variation in the
tropical Pacific air-sea coupled system, occurring at
irregular intervals of 2 to 7 years and lasting 9 months to 2
years\cite{Cane83}; it is well known through its linkage
to natural disasters in far flung regions of the globe, such as the
floods in Ecuador, the droughts in Southeast Asia and Southern Africa, the
increased number of storms over the Pacific Ocean. There are several indices
measuring the strength of \ENSO, the popular ones being \Nino3 and
\Nino4\cite{Trenberth97}. 
IOD is another air-sea coupled climate mode, characterized by
an aperiodic oscillation of sea surface temperature (SST) 
in the Indian Ocean\cite{Saji99}.
It has been related to, among others, the floods in East Africa and 
drought in Indonesia and parts of Australia. IOD is measured by an index
called Dipole Mode Index (DMI).

As the dominant modes in respectively Pacific and Indian Oceans, 
the relationship between \ENSO and IOD is of substantial interest;
see \cite{Wang04} for an excellent review.
It has long been recognized that the tropical Indian and Pacific 
Oceans are interrelated\cite{Kiladis} 
on their SST anomalies. But the relation between the two modes
is still to be clarified. In general, it is believed
that \ENSO may induce IOD, which usually peaks in fall; a recent
study is referred to \cite{Wang13}.
The causality of \ENSO is easily
understandable, considering the alteration of the Walker circulation during
the \ENSO years which causes the subsidence over the Indian Ocean 
(e.g., \cite{Klein99}).
On the other hand, the impact from the IOD has jut been recognized
recently; in fact, in early efforts, Indian Ocean used to be 
treated as a slab of mixed layer responding passively to \ENSO\cite{Lau00}.
The recognition is mainly
through climate predictions, such as the \ENSO predictions 
(e.g., \cite{Chen08}, \cite{Izumo14}) 
and the CGCM experiments (e.g., \cite{Annamalai10}).  
Due to the correlation between the two modes, it has
been suggested that an Indo-Pacific perspective should be adopted in
researching on the \ENSO and IOD  problems (e.g., \cite{Webster10})

The linkage between IOD and \ENSO, though evidenced from different
aspects, is somewhat still an issue in debate. This is partly due to the
failure to recover the IOD pattern in the Indian Ocean in many correlation 
or time-lagged correlation analyses with the indices and the SST
observations, albeit the correlation could be significant.
Such an example is shown in the Fig.4 of Wang et al. (2004).
This has led people to conclude with caution that IOD might be 
partially independent of \ENSO, though \ENSO tends to induce IOD. 
This problem, which is obviously a problem on cause and effect,
could be due to the inadequacy of using correlation analysis for causality
purposes. Now that we have arrived at the formula (\ref{eq:caus21}), 
let us see how things will come out when it is applied to 
the IOD-\ENSO causality analysis.

First look at the causal relation between the indices of the two modes.
The DMI used is the monthly series by Japan Agency for Marine-Earth 
Science and Technology (JAMSTEC), which can be downloaded from their website. 
The \ENSO indices are from NOAA ESRL Physical Sciences Division
(http://www.esrl.noaa.gov/psd/); they are also monthly data.  
Since this DMI series spans from January 1958 to September 2010, 
the \ENSO indices, which have a much longer time span (1870-present), 
are also tailored to the same period. These series then have a total of
633 time points. Application of (\ref{eq:caus21}) to the DMI and \Nino4
yields a flow of information from \ENSO to IOD: $T_{E\to I} = -6$, and 
a flow from IOD to \ENSO: $T_{I\to E} = 13$ (units: $10^{-3}$ nats/month).
Using \Nino3 one arrives at a similar result: 
$T_{E\to I} = -6$ and $T_{I\to E}=16$.
That is to say, \ENSO and IOD are mutually causal, and the causality is
asymmetric, with the one from the latter to the former larger than its
counterpart. Moreover, the different signs indicate that \ENSO 
tends to stabilize IOD, while IOD tends to make \ENSO more uncertain.

By our previous experience, series length is crucial.
To see whether the length here is adequate, we have tried 
shortened series: 1963-2010, 1968-2010. The former yields essential the
same result as the one in its full length (1958-2010); the result of the 
latter is also similar. In fact, even series with a span as short as
1970-2010 can give qualitatively similar result. So the series may be long
enough to form an ensemble with sufficient statistics. The resulting 
information flow rates hence may be acceptable.

\begin{figure}[h]
	\begin{center}
	\includegraphics[angle=0, width=1\textwidth]
		{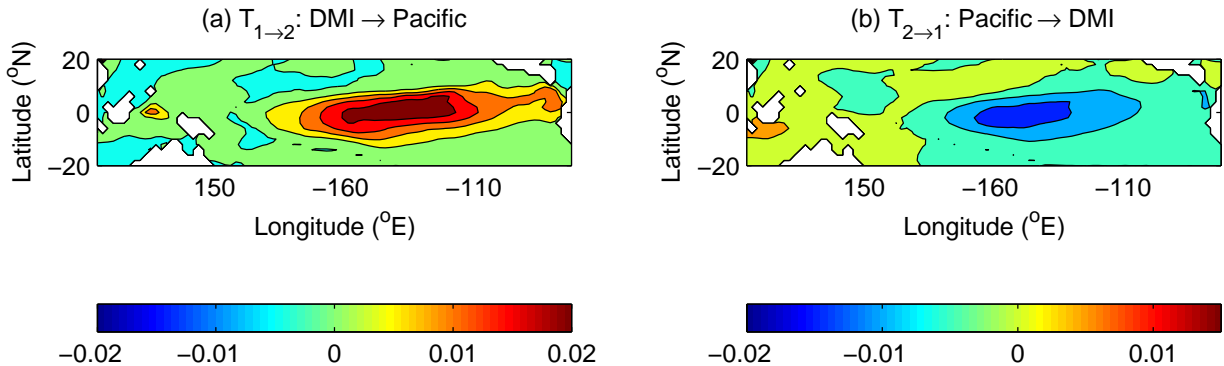}
	\caption{Information flow between DMI and the tropical Pacific SST
	(in nats/month).
	 \protect{\label{fig:Pacific}}}
	\end{center}
	\end{figure}

With this assertion we proceed to extract the causality patterns out
of the SST in the Pacific and Indian Oceans. The SST data are from the
above NOAA site. As before, we use only the data for the period 1958-2010.
First, compute the causality between DMI and the tropical Pacific SST.
The results are shown in Fig.~\ref{fig:Pacific}. 
Clearly DMI and the Pacific SST are mutually causal, 
and the information flow in either direction has a \ENSO-like pattern.
The signs of the flows are the same as computed above with two index series.

\begin{figure}[h]
	\begin{center}
	\includegraphics[angle=0, width=1\textwidth]
		{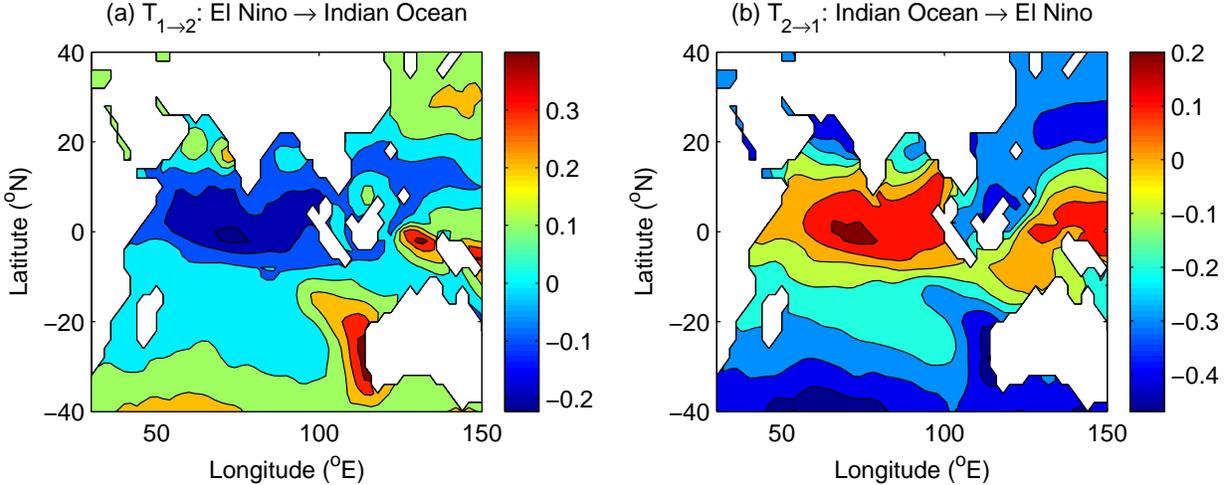}
	\caption{Information flow between \Nino4 and the Indian SST
	(units: nats/month).
	 \protect{\label{fig:Indian}}}
	\end{center}
	\end{figure}

If the above \ENSO-like pattern in the Pacific is common, 
the following pattern in the Indian Ocean is unique to our 
causality analysis. Compute the information 
flow between \Nino4 and the Indian Ocean SST, and plot the result in
Fig.~\ref{fig:Indian}. Fig.~\ref{fig:Indian}a shows the flow from \ENSO
to the SST. In the tropical region, there are indeed two poles, though the
eastern one covers a rather limited region over Indonesia.
On the map of the feedback from the Indian Ocean SST
(Fig.~\ref{fig:Indian}b), this structure is much clear, with a dipole
pattern reminiscent of IOD, though there seems to be a shift eastward. 
This is remarkable,
as in previous correlation analyses or time-lagged correlation analyses,
the IOD pattern is not seen. Note in Fig.~\ref{fig:Indian}b, both centers
are positive, indicating that the Indian Ocean influences the \ENSO through
IOD, which functions to amplify the \ENSO oscillations. In contrast, the
impact of \ENSO on the Indian Ocean SST divides over the two centers, which
have opposite signs of information flow.
The causality analysis with \Nino3 shows similar patterns,
though the features may not be as pronounced as that with \Nino4.

The relation between \ENSO and IOD is an extensively investigated subject;
further discussion, however, is beyond the scope of this study. Our purpose 
here is to use it as one real world example to demonstrate the utility of 
the newly derived formula (\ref{eq:caus21}). 
The result is encouraging: the IOD-like pattern, which is anticipated as
evidence builds up, but missed in previous correlation studies, 
is recovered faithfully in the index-SST causality patterns. 
Moreover, it is found that, on the whole, the \ENSO influences the 
IOD by making the latter more certain, while the causality from IOD to 
\ENSO is generally the opposite. This is perhaps the reason why 
it has long been observed in a definite way that \ENSO may induce IOD,
while the influence from IOD to \ENSO, albeit stronger, has been just 
recognized recently
through predictability studies. The latter, which carries a positive
information of flow, means that, to \ENSO, the Indian Ocean is a source of
uncertainty, and the causality from IOD to \ENSO is manifested as a
transference of uncertainty from the former to the latter. Clearly
an efficient way to observe and forecast an event is to put extensive 
observation at its uncertainty source--a way how target observing systems 
are designed.  This is why knowledge of the Indian Ocean
facilitates the \ENSO forecast (e.g., \cite{Chen08}, \cite{Izumo14}),
i.e., increases the predictability of \ENSO.

\section{Discussion}	\label{sect:discussion}

We have obtained, based on a rigorous formalism of information flow, 
a concise formula 
for causality analysis between time series. For series $X_1$ and
$X_2$, the rate of information flowing (units: nats per unit time) 
from the latter to the former is
	\begin{eqnarray}
	\info_{2\to1} = \frac {C_{12}} {C_{11}} \cdot 
			\frac {C_{11} C_{2,d1}-C_{12} C_{1,d1}} 
			      {C_{11}C_{22} - C_{12}^2}, 
	\end{eqnarray}
where $C_{ij}$ is the sample covariance between $X_i$ and $X_j$, 
$C_{i,dj}$ the covariance between $X_i$ and $\dot X_j$,
and $\dot X_j$ the difference approximation of $\dt {X_j}$ 
using the Euler forward scheme [see Eq.~(\ref{eq:xdot})]. 
If $T_{2\to1} = 0$, $X_2$ does not cause $X_1$; if not, it is causal.
In the presence of causality, two cases can be distinguished according to
the sign of the flow: a positive $T_{2\to1}$ means that $X_2$ functions to
make $X_1$ more uncertain, while a negative value indicates that $X_2$
tends to stabilize $X_1$.
The formula is very tight, involving only the common statistics namely
the sample covariances. It has been validated with touchstone series, where 
the stationary preset one-way causality can be rather accurately recovered, 
provided that the series is long enough 
(time span, not number of data points)
to contain sufficient statistics. 
When the series length is guaranteed, the formula works even with series 
with rather coarse resolution. Moreover, the series need
not be stationary for the formula to apply; in fact, when a nonstationary
segment is included, it works with much shorter series. In this case,
the equation should be replaced by
	\begin{eqnarray}
	\info_{2\to1} = \frac {C_{12}^*} {C_{11}^*} \cdot 
			\frac {C_{11} C_{2,d1}-C_{12} C_{1,d1}} 
			      {C_{11}C_{22} - C_{12}^2}, 
	\end{eqnarray}
where $C_{ij}^*$ are the sample covariances computed using the slab of 
data with nonstationarity removed.
In practice, we suggest that one select out the stationary part of the data, 
or detrend the series, before estimating $C_{ij}^*$. But when computing
$C_{ij}$, the original data should be used; no detrending or other manipulation
should be made.

It should be pointed out that this formula is based on a
formalism with respect to Shannon entropy, or absolute entropy, 
while it has been
argued\cite{Kleeman02}, in predictability research, relative entropy is a 
more advantageous choice due to its nice properties such as invariance 
under nonlinear transformation. Fortunately, as we have proved in 
\cite{Liang13_physica}, for 2D systems, the information flow thus obtained 
is the same with both absolute and relative entropies. 
But, of course, the sign has to be changed, as the former is for
uncertainty, and the latter for predictability.

Some issues remain. Firstly, what we have arrived is for linear systems,
while in reality, nonlinearity is ubiquitous. For sure there is still a 
long way to go along this line of research, and this study just marks
the starting point. Secondly, the time span of the series should be long
enough for the estimation to be accurate. But how long? It would be useful
if we can have some criterion to judge whether a given length is adequate
or not. Thirdly, the rates of information flow differ from 
case to case; one would like to normalize them in real applications. 
A natural normalizer that comes to mind, at the hint of correlation
coefficient, might be the information of a series 
transferred from itself. The snag is, this quantity may turn out to be 
zero, just as that in the H\'enon map, a benchmark problem we have 
examined before (cf.~\cite{Liang13}). 
These issues, among others, will be considered in the forthcoming studies.

\section*{Acknowlegments}
This study was partially supported by Jiangsu Provincial Government through
the Jiangsu Chair Professorship to XSL, and by
the National Science Foundation of China (NSFC) under Grant No.~41276032.

\end{document}